IAC-22- D2.7.7.x68906

# Development, manufacturing and testing of small launcher structures from Portugal

**Andre G.C. Guerra[a]*, Daniel Alonso[b], Catarina Silva[b], Alexander Costa[a], Joaquim Rocha[a], Luis Colaço[a], Sandra Fortuna[a], Tiago Pires[a], Luis Pinheiro[a], Nuno Carneiro[a], André João[a], Gonçalo Araújo[b], Pedro Meireles[b], Stephan Schmid[b]**

[a] CEiiA *Av. Dom Afonso Henriques, 1825. 4450-017 Matosinhos, Portugal*, magellan@ceiia.com
[b] RFA-Portugal, Av. Dom Afonso Henriques, 1825. 4450-017 Matosinhos, Portugal, catarina.silva@rfa.space
* Corresponding Author

**Abstract**

During the last decades the Aerospace Industry has seen the number of Earth orbiting satellites rise at a stunning rate. This race stems from the need to monitor Earth and better understand its environments at different scales as well as to establish global communication networks, for example. Nano, micro, and small satellites have been a prime tool for answering these needs, with large and mega constellations planned for the near future, leading to a potential launch gap that can only be answered by an increase on the number of yearly launches, to keep up with demand as well as replenish established capacity. An effective and commercially appealing solution is the development of small launchers. These can complement the current available launch opportunity offer, serving a large pool of different types of clients, with a flexible and custom service that large conventional launchers cannot adequately assure. Rocket Factory Augsburg, who has been developing its own small launcher for the last two years, has partnered with CEiiA, a Portuguese engineering and product development centre, for the development of several structures for the RFA One rocket. The objective has been the design of solutions that are low-cost, light, and custom-made, applying design and manufacturing concepts as well as technologies from other industries, like the aeronautical and automotive, to the aerospace one. This allows for the implementation of a New Space approach to the launcher segment, while also building a supply chain and a set of solutions that enables the industrialisation of such structures for this and future small launchers. The two main systems under development have been a versatile Kick-Stage, for payload carrying and orbit insertion, and a sturdy Payload Fairing, both with multiple configurations. Even though the use of components off-the-shelf have been widely accepted in the space industry for satellites, these two systems pose different challenges as they must be: highly reliable during the most extreme conditions imposed by the launch, so that they can be considered safe to launch all types of payloads; while allowing for the maximum payload mass and volume, making its business case sustainable. Additionally, the manufacturing methods had to be such to allow the easy scale up and ultimately the creation of a production line for these structures. This paper thus dives deep on the solutions developed in the last few years, presenting also lessons learned during the manufacturing and testing of these structures.

**Acronyms/Abbreviations**

| | | | |
|---|---|---|---|
| PF | Payload Fairing | SIF | Separation Interface Flanges |
| KST | Kick-Stage | RPN | Risk Priority Number |
| S | Severity | FMEA | Failure Mode and Effect Analysis |
| O | Occurrence | RFA | Rocket Factory Augsburg |
| P | Probability | ESPA | Expendable Secondary Payload Adapter |
| D | Detection | LV | Launch Vehicle |

## 1. INTRODUCTION

### *1.1. MAGELLAN PROJECT*

The projected number of satellites to be launched has been increasing at a substantial rate, with a significant portion of those future satellites as part of large constellations, in some cases ascending to the hundreds of elements, mainly from the commercial sector (over 60%) and linked to Earth observation (45%) and communications (19%) [1]. These missions, besides the number of planned satellites, require an increased launcher flexibility or even dedicated launches, so not to incur in loss of revenue or opportunities.

Several projects for micro-launcher development or adaptation of large launcher capacity have been initiated around the globe to answer this clear market need for increased launcher capabilities. In Europe there are several initiatives like the one from the European Space Agency (ESA) for Light satellites Low-cost Launch service [2] or the one from Rocket Factory Augsburg (RFA) with their RFA ONE rocket.

Portugal has a unique and important opportunity in this expanding strategic market. The growing Portuguese installed competences, in alignment with the Portuguese space strategy [3], allows Magellan to bring together two forward looking entities with complementary knowledge. RFA-PT, a subsidiary of Rocket Factory Augsburg, with the Aerospace know-how and global industry recognition, and CEiiA, an engineering and product development centre with a long track record of developing disruptive solutions in many different markets.

Magellan aims at developing, prototyping and testing a set of primary structures for micro-launchers, namely the Payload Fairing (PF), protecting the payload from flight loads, the Kick-Stage (KST), a versatile satellite like structure to support all payloads and carry them to their final orbits, and the Separation Interface Flanges (SIF), which connect the different stages of the vehicle. A photo of the engineering models prepared so far is shown in Fig. 1, together with part of the development team both from RFA and CEiiA.

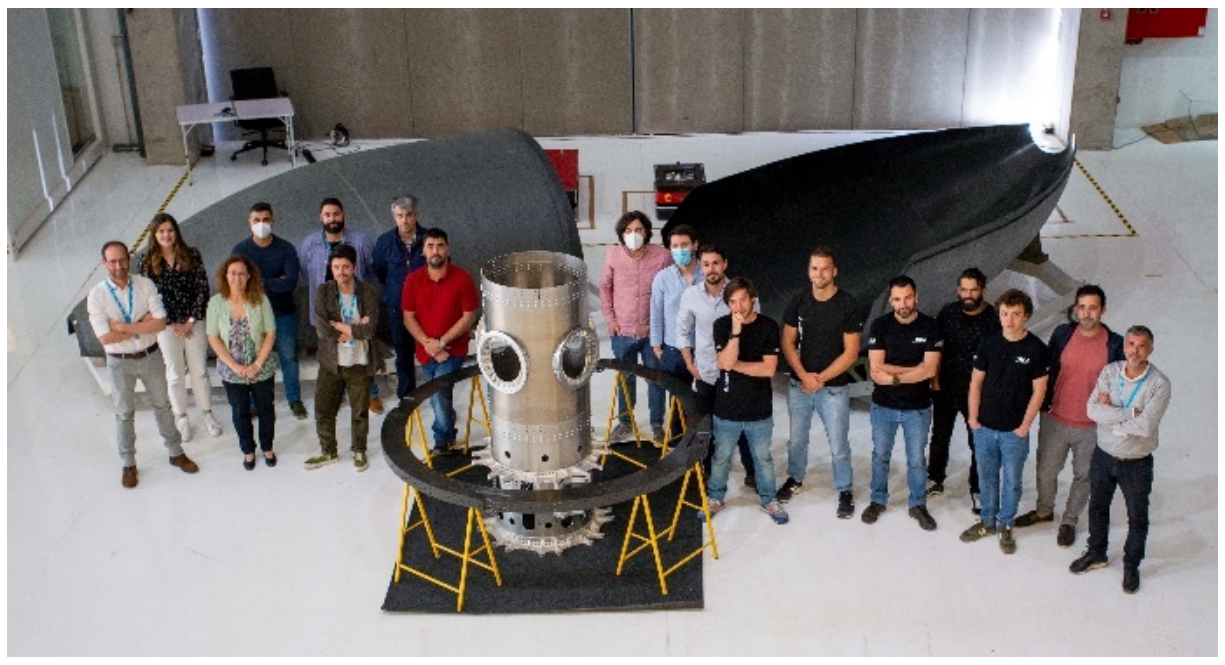

Fig. 1. Project team and structural components under development

Although based on a RFA's specific case, the developments undertaken could be transposed to any other future actors in the space transportation business, as well as on other sectors including the satellite business and the aeronautical one.

As such, RFA-PT and CEiiA are developing together significant micro launcher structures in Portugal, within this ambitious Magellan project, and demonstrate the Portuguese potential to develop complex, low-cost, light and custom-made aerospace solutions from concept up to manufacturing in an enduring mutually beneficial partnership with global industry partners at scale.

### *1.2. RFA ONE – DETAILS AND SPECIFICATIONS*

Rocket Factory Augsburg was founded in 2018 with the vision to enable data generating business models in space to better monitor, protect and connect our planet Earth. Against this background, the company's goal is to offer launch services of up to 1 300 kg into low Earth orbits, and beyond, on a weekly basis at unmatched prices. With this, RFA wants to democratise access to space and reduce the launch costs in the space industry. The RFA ONE launch service combines three key competitive advantages: A customer focused service with precise in-orbit delivery and a high degree of mission flexibility through its orbital stage; at a highly competitive price; made possible by superior staged combustion technology, low-cost structures and usage of industrial components.

RFA ONE is a three-stage expendable launch vehicle (Fig. 2), powered by RFA's in-house developed kerolox ORSC "Helix" engine. The vehicle is constructed from inexpensive stainless-steel featuring common bulkhead tanks and automotive-grade composite inter-stage skirts. Later variants of the launch vehicle will allow for first stage recovery and engine reusability. Its Kick-Stage is the brain of RFA ONE and is more akin to a satellite than a traditional rocket stage. During flight, it steers and monitors the whole vehicle, ensuring navigation, data acquisition, and telemetry, as well as throttle management and jettisoning sequences. The configurable payload fairing volume allows for various payload configurations. Once in orbit, the orbital stage is a fully functioning autonomous spacecraft. It is able to dispatch multiple payloads performing several manoeuvres, including altitude and inclination changes. This offers unmatched payload delivery capabilities and mission flexibility after launch. As an added feature, the orbital stage is designed for a lifetime exceeding 12 months with reliable de-orbiting systems compliant with the international standards for the space debris mitigation.

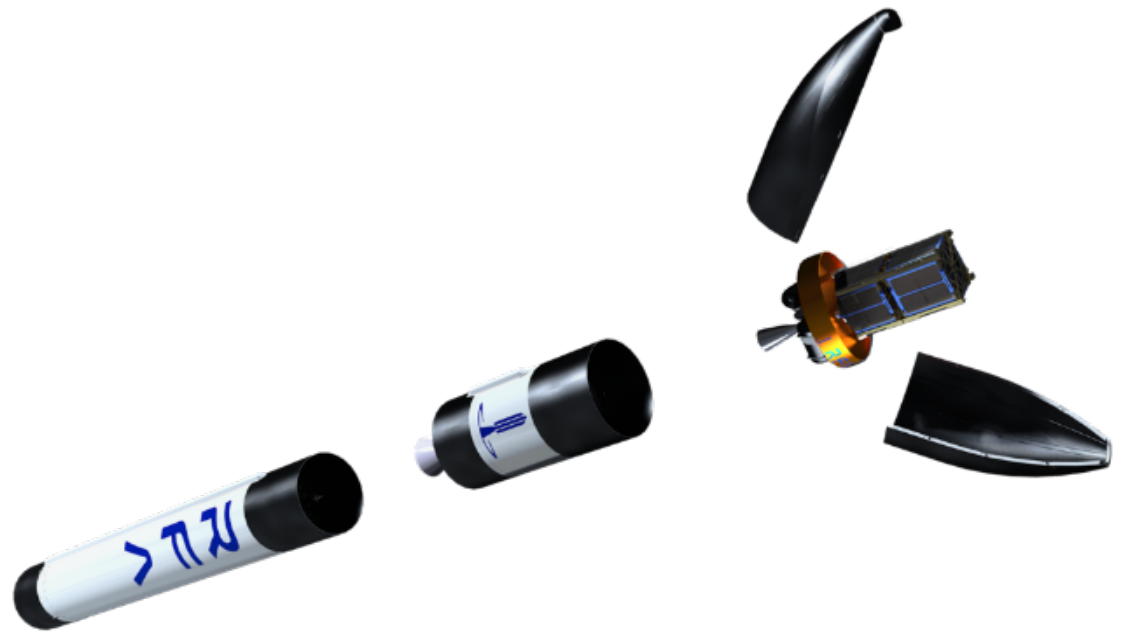

Fig. 2. RFA ONE vehicle stages

Due to its versatility of design for both the Payload Fairing and the Kick-Stage, there are several payload configurations available: dedicated launch for one single customer; or rideshare configuration for several customers.

RFA ONE will make its debut flight in 2023.

## 2. DEVELOPMENT METHODOLOGIES

### *2.1. SYSTEMS ENGINEERING*

Due to the complexity of developing both the KST and PF and the challenges imposed by the different disciplines involved throughout the product development cycle, a Systems Engineering approach has been implemented in Magellan.

This approach is often used in the development of highly complex systems, with several stakeholders and with multi-disciplinary activities that frequently share conflicting constraints. In fact, NASA has early identified the benefits of Systems Engineering applied to Space projects [4], defining it as *"a methodical, multi-disciplinary approach for the design, realization, technical management, operations, and retirement of a system"*. The first organisation to codify a formal systems engineering process was the United States DoD captured in MIL-STD-499 [5], which was cancelled in 1995 and later replaced by recent commercial standards, of which the most frequently used are IEEE-1220 [6], the EIA-STD-632 [7], and the ISO-IEC-IEEE-STD-15288 [8].

A top-level diagram of the proposed Systems Engineering approach for Magellan, using as a starting point the previous standards, as defined in [9], is depicted in Fig. 3.

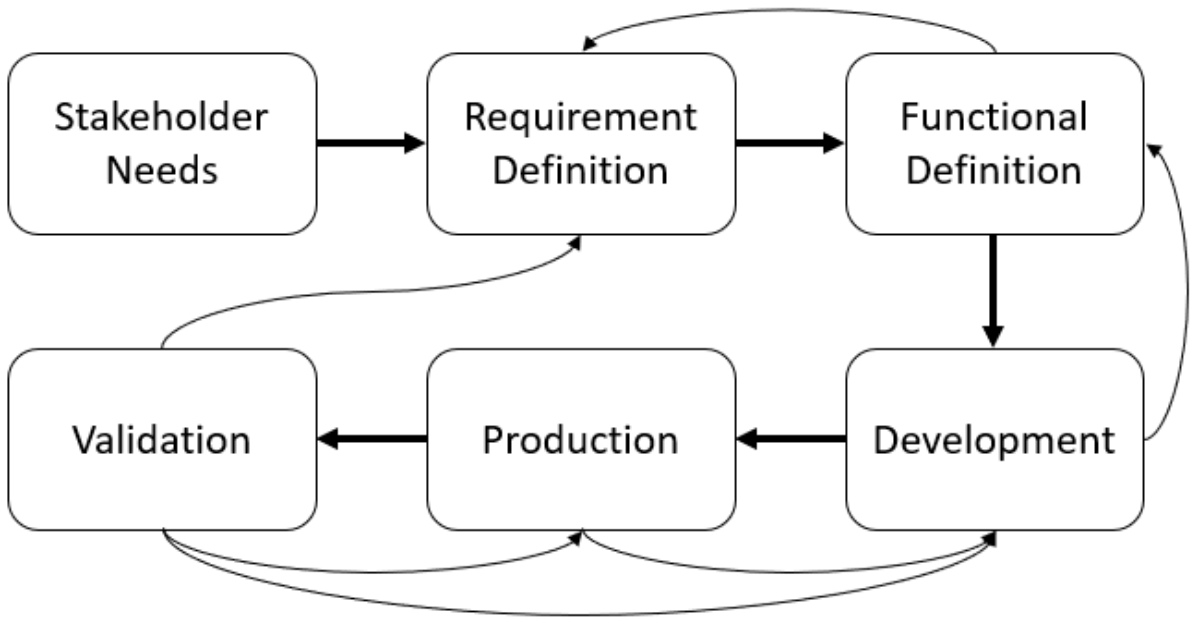

Fig. 3. Proposed Systems Engineering top-level diagram

The main stages of this approach include:

- **Requirement Definition:** in this stage, all the input conditions including requirements, plans, milestones and other relevant are assembled and organised. The KST/PF operational needs and expected performance are defined to establish the requirement constraints.
- **Functional Definition:** this stage translates requirements into functions (actions and tasks) that the system and sub-systems must accomplish, defining the interactions among functional elements to lay a basis for their organisation.
- **Development:** during this phase, several conceptual trade-off designs are evaluated, converging into a selected solution by analysing a set of predefined and prioritised criteria. The solution is then detailed by iterative concurrent engineering approaches between all involved disciplines, culminating in relevant and necessary information for the solution materialisation. During this stage, several iterations of structural Design and Analysis for KS, SIF and PF took place. The information generated were 2D and 3D CAD models, a complete Bill of Materials and a Manufacturing Specifications Document.
- **Production:** with the generated Engineering specifications, and after its approval, the products proceed to the manufacturing and assembly stage. Before delivery, a verification of the executed parts and assemblies, against the manufacturing specifications, is performed.
- **Validation:** in this last stage, the final solutions of the KST, SIF and PF are evaluated against the requirements, both at system and, when needed, sub-system level. This validation confirms that the solutions perform as expected, allowing any correction if necessary, and thus may occur during any stage of the development, through simulation, testing, inspection, or any other method identified during requirements definition.

Regarding the Systems Engineering model implemented, it is worth mentioning that this approach is dynamic, versatile and iterative throughout the complete process, allowing to go back and revisit previous stages if any management or technical constraint is verified.

Finally, and regarding the process depicted in Fig. 3, transversal Systems Engineering activities are taking place during the entire project. More specifically, Technical Risk Analysis, Failure Mode and Effect Analysis, Technical Planning, Interface Management, scheduling and Technical Revies are examples of such activities implemented.

### *2.2. REQUIREMENTS DEFINITION & MANAGEMENT*

System requirements are an approach to meet mission objectives paving the way for the development of the system. The definition of requirements for PF and KST subsystem rely primarily on the fundamental Launch Vehicle (LV) objectives, which provide the Top-level requirements. These then breakdown into three basic types of requirements: functional, operational and constraints.

In the requirements definition for these systems, care was taken in the way of defining them. The requirements specify the goal that is required to achieve, without specifying how to achieve it. These let the systems engineering trade process to be able to explore multiple alternatives and realize trade-offs to identify the one which provides improving performance at minimal increase of cost/risk.

The LV primary objective is to transport payload into specific orbits, which translates into top level requirements of mass and volume. These are among the critical key performance requirements. From flight and mission conditions more functional and operational requirements are determined and also project constraints. These then propagate to each system and subsystem, which then will reply with a set of constraints that each (sub)system will define.

The intention is for the requirements to evolve to achieve the objectives at minimum cost/risk. Concentrating system trades around critical requirements of each system, by constantly checking and re-checking the ones that highly impact on performance, cost and risk of the system.

### *2.3. RISK ANALYSIS & FMEA*

As previously stated in Section 2.1, various Systems Engineering activities were performed during the complete project course. In this section, special focus will be given to Risk Analysis and Failure Mode and Effect Analysis (FMEA) as these methods have highly contributed to accelerate the product development cycle and quickly converge to a final functional solution/ prototype.

#### *2.3.1. Risk Analysis*

The implemented risk analysis plan considers the following stages:

- **Risk identification:** a risk ID and opening date are assigned and a comprehensive description is detailed.
- **Risk measurement:** the risk exposure is calculated by multiplying the risk Probability (P) by the risk Severity (S). This is a qualitative analysis performed by the entire technical team with different background and experiences. Probability and severity classification follow the specifications in Table 1 and Table 2.

Table 1. Risk probability classification

| ***Probability (P)*** | **Description** |
|---|---|
| *1* | Low (< 20%) – Unlikely to occur |
| *2* | Medium Low (20 – 50%) – Low probability of occurring |
| *3* | Medium High (50 – 80%) – Large probability of occurring |
| *4* | High (> 80%) – Very likely to occur |

Table 2. Risk severity classification

| ***Severity (S)*** | **Description** |
|---|---|
| *1* | No impact or no significant impact. |
| *2* | Significant impact or minor project effects but no critical impact. |
| *3* | Serious impact with significant effects on the project. Possible effects on the final result. |
| *4* | Severe impact on the project and final results. Noncompliance with the requirements. |

After Probability and Severity classification, the Risk Exposure is calculated ($RE = P \times S$). For each risk, the value is compared to the risk exposure matrix in Table 3.

Table 3. Risk Exposure Matrix

| | Risk Exposure | | | |
|---|---|---|---|---|
| | Probability x Severity | | | |
| 4 | *4* | *8* | *12* | *16* |
| 3 | *3* | *6* | *9* | *12* |
| 2 | *2* | *4* | *6* | *8* |
| 1 | *1* | *2* | *3* | *4* |
| | 1 | 2 | 3 | 4 |

Based on the calculated RE, a different strategy has been followed, namely:

- **Mitigation: i**n Magellan, mitigation plans were implemented for RE >8, and is used to reduce risk probability, severity, or both.
- **Contingency Plan:** defined for RE >12, allow the specification of the required steps to be followed when a risk is verified, to reduce its impact.

Risk analysis is being implemented throughout the complete project duration, with risks revisited and updated (if required) every 3 months. These include technical situations as well as management aspects, such as components lead time, human and machinery resources availability and other.

#### *2.3.2. FMEA*

Failure Mode and Effect Analysis is a systematic method for evaluating a process to identify possible failure points and their impact in the overall process performance [10]. FMEA can be as complex as the implementing team desires, so it must be a well-balanced method between the effectiveness and the available effort capacity from the development team.

As risks are defined as hazardous situations that may impact the project, product, users, or operational environment, their analysis can be complemented by a well-established FMEA. In fact, risks can (but not only) be originated from failures occurring anytime during the product lifecycle, leading to a specific effect which, if not timely identified, may represent a catastrophic impact.

Similarly, to the risk analysis in Section 2.3.1, FMEA is built using three main variables:

- **Severity (S):** classification (1-10) of the failure impact in the process/product.
- **Occurrence (O):** classification (1-10) of the failure probability.

- **Detection (D):** classification (1-10) of the capacity to detect the failure through the imposed analysis (maintenance, inspections, testing, etc).

After identifying and classifying failures, the Risk Priority Number (RPN) is calculated as $RPN = S \times O \times D$.

The RPN allows the overall classification of the failure. However, is advisable to rely in more than the RPN, when classifying critical systems, since failures may have different Severities in the system, for the same RPN.

This method was applied to the KST, SIF and PF products development process, from component level up to subsystem and system, when considered essential. Moreover, a simultaneous comparison between Severity and RPN is performed, following the matrix in Fig. 4 [11].

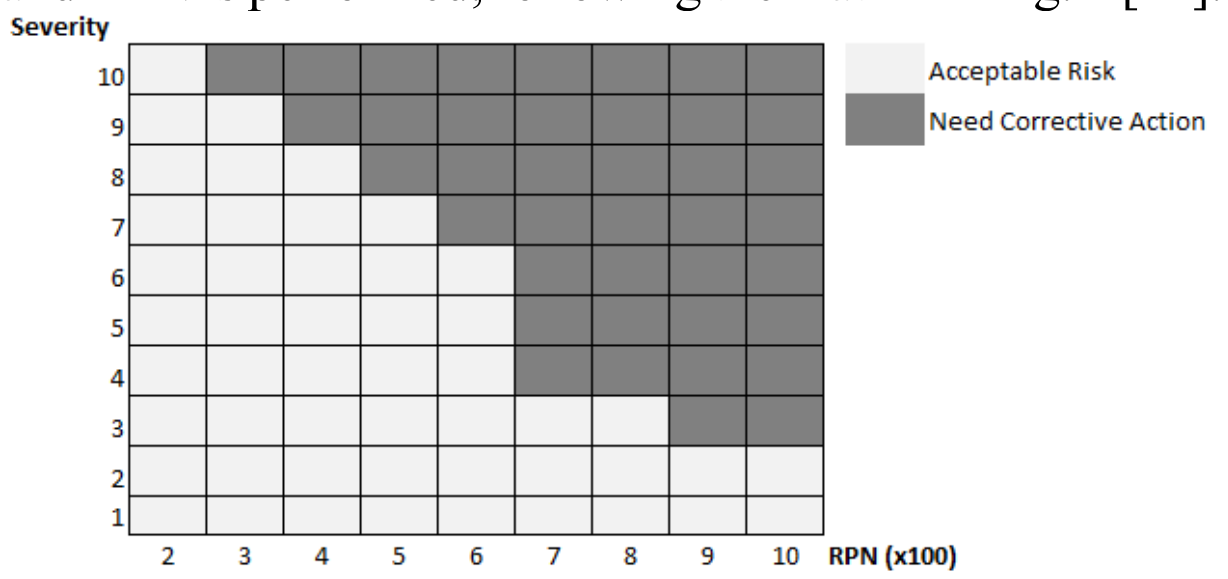

Fig. 4. Severity vs RPN matrix implemented

The RPN matrix in Fig. 4 was customised by the project team. The cells highlighted in dark tone, represent a combination of values that impose the implementation of a corrective action. This action needs to be detailed and have a direct impact in at least one of the variables that calculate RPN. The new value must then be updated and the result to bring the failure to the light shaded area. This action needs to be registered apart from the initial evaluation, to keep the entire FMEA traceability. Similarly, to the risk analysis, FMEA is revisited and updated (if required) every 3 months.

The implementation of Risk Analysis and FMEA in Magellan, allowed to increase the efficiency of the development process by constant monitoring risk and failure points. This implementation ultimately led to time and cost optimisation and increased reliability throughout the expected product lifecycle.

## 3. DEVELOPMENT

### *3.1. DESIGN*

The Design main driver was the lightweight and low-cost materials that can be found in several transportation industries, such as aeronautical and automotive ones. The use of composite materials was highly recommended for its low weight to strength ratio, as well as sheet metal working due to its simplicity and low cost of manufacturing. Machined parts were developed as a last resort when the previous two technologies become too complex or expensive to manufacture.

#### *3.1.1. Payload Fairing (PF)*

The PF is a bullet shaped shell, divided in three sections. The left-hand and right-hand shells are composed of a composite skin, reinforced with flanges to provide an interface between each half. On the tip of the fairing a separated dome, later attached to one of the shells, closes the shape of the PF.

With modularity in mind the composite and the flanges were designed in such a way to allow for multiple PF sizes, ranging from 4 meters in length up to 6 meters, reducing costs on manufacturing by re-using already produced parts.

A concept schematic of the PF is depicted in Fig. 5.

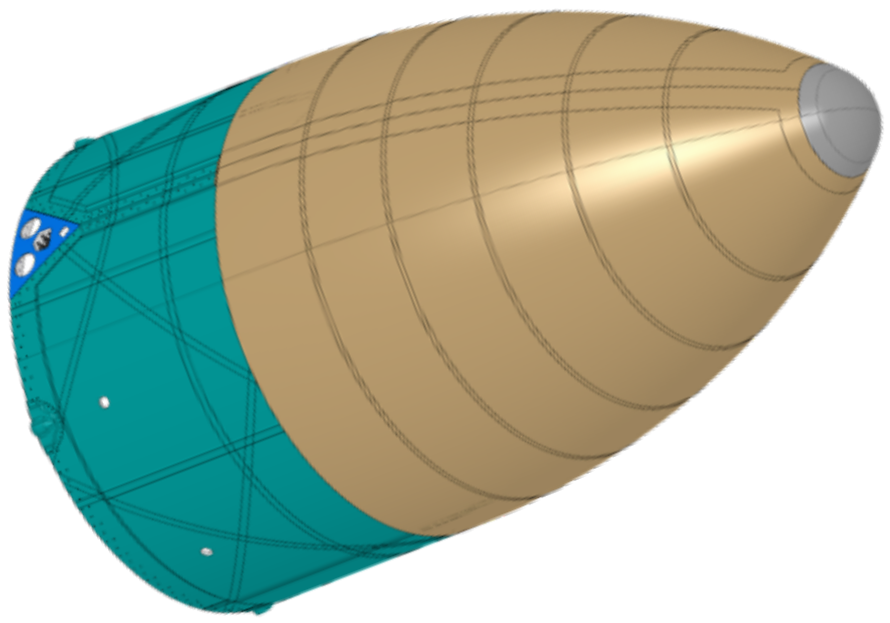

Fig. 5. Payload Fairing general concept

#### *3.1.2. Kick-Stage (KST)*

With a multitude of missions and modularity in mind, the KST design was based on an Expendable Secondary Payload Adapter (ESPA) concept tower.

Making use of automotive technologies and materials, the KST evolved into a cylindrical shaped vehicle, with three configurations possible. The first configuration, designated ESPA-0, comprised a base cylinder, housing all the necessary systems for operation, and a payload adapter, compatible with commercially available satellite lightbands. The ESPA-1 and ESPA-2 (Fig. 6) configurations see an additional one or two payload cylinders added, respectively, in between the base cylinder and the payload adapter.

This orbital transfer vehicle has, structurally, 1 meter in diameter and rises from 1.15 meters to 3 meters, depending on the configuration used (i.e. the number of payloads to transport).

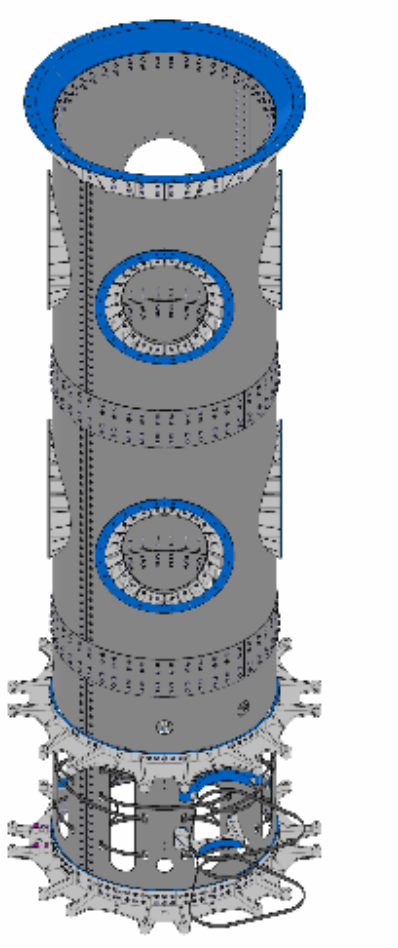

Fig. 6 Kick-Stage ESPA-2 configuration structures

### *3.2. STRUCTURAL ANALYSIS*

The approach for the structural analysis is based on FEA with sufficient detail to evaluate the failure in composite layups, metallic components and fasteners (Fig. 7). The detail includes multiple parts and non-linear analysis to provide correct load paths using contacts. This allows for a time saving in the definition of substructures and a correct interdependence of the modification of each structure on the surrounding components.

Static and dynamic load cases were defined from the requirements and launch loading events, with consideration for the factors of safety established in the requirements. Inertial and aerodynamic pressure loads were applied to a finite element representation of the modelled structures (Fig. 8), elaborated from CAD information, in accordance with industry practices.

The results were evaluated for the components and structures directly from the FEA, whilst the fasteners were subjected to evaluation with empirical methods defined by ECSS standards.

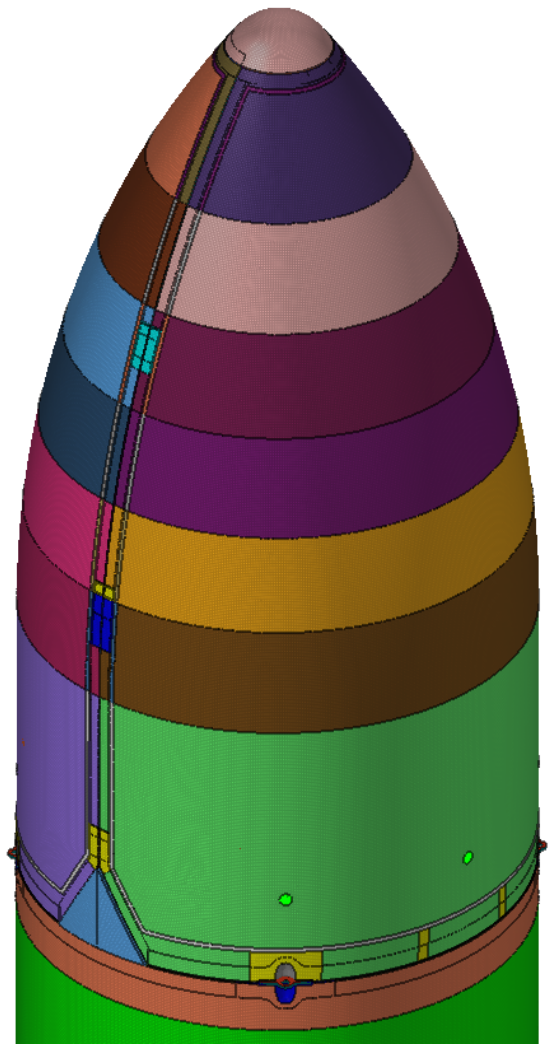

Fig. 7. Payload Fairing FEM model

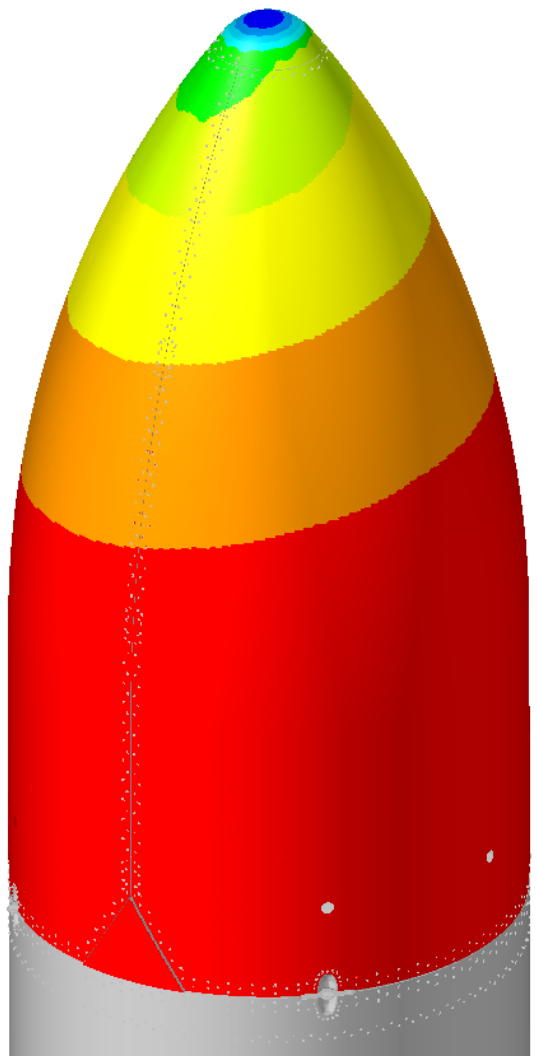

Fig. 8. Aerodynamic pressure distribution on the Payload Fairing

### *3.3. PRODUCTION & ASSEMBLY*

The manufacturing of the components that are part of the Magellan project presented significant challenges, both on the production processes and methodologies that had to be followed. These must be based on three crucial cornerstones: low-cost end products, lean and fast timelines, and high-quality components. To ensure these three cornerstones, several studies were carried out, including, for instance, on composite materials, where the behaviour of the materials during curing cycles, in terms of deformations, was analysed to adapt the mould and the part shape and configuration. These studies helped to select production processes that allowed the production of low-cost parts while maintaining their superior quality.

The previous know-how of the automotive and aeronautical industry combined with aerospace processes and methodologies were also essential to ensure the quality in all the manufacturing and assembly process. These methodologies were used for assembling several components (as shown in Fig. 9) and were paramount to guarantee a balance between quality and time throughout the full process.

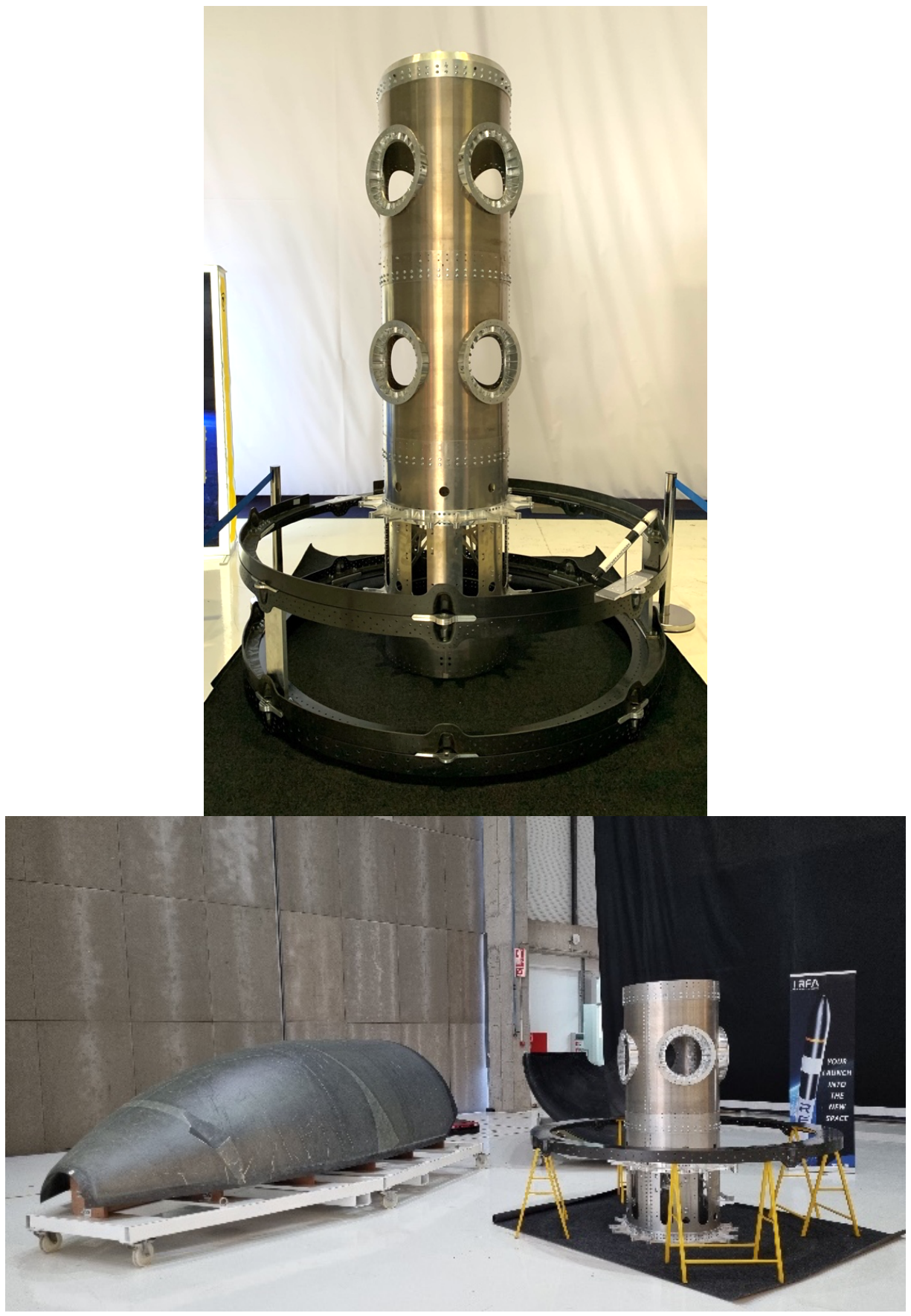

Fig. 9. Kick-Stage and Payload Fairing engineering models

## 4. TESTS

### *4.1. TEST PHILOSOPHY*

The PF and KST structures test campaign, performed by RFA PT, is designed to provide sustaining proof that these structures are flight worthy and, therefore, able to protect and deliver the payload to its desired orbit without any damage.

The dimensioning rules and qualifications processes of the different structures have distinct goals and functionalities, which leads to differently designed test campaigns. Each system has its own tailored tests designed to better suit and provide valuable information and proof of concept.

The PF static test campaign provides relevant information for the fairing design and validation, allowing to assess:

- The PF capability to survive and maintain performance after being subjected to the quasi-static loads foreseen for the worst condition during launch;
- FEM correlation and validation;
- Gain knowledge about the structure's behaviour and its limits.

The main purpose of the KST test campaign is to evaluate its structural integrity and transmissivity to the payload, as well its subsystems, namely through:

- Behaviour of the KST structures to flight loads,
- Intrinsic properties – resonant frequencies, damping characteristics, displacements;
- FEM correlation and validation;
- Gain knowledge about the structure's behaviour and its limits.

The testing philosophy relays on mathematical models to justify test configurations and provide correction factors to cover different effects relative to flight case. Usually, the dimensioning flight case is a very complex load case originated by simultaneous loads (static, dynamic, aerodynamic, temperature) and very difficult to reproduce on a test stand. Test configurations overcome these difficulties by applying simplified loading conditions when compared to the dimensioning load case conditions on a flight similar structure. These are designed to induce similar levels of solicitations at dedicated locations, supported by mathematical models. Once the specified stress/strain fields are achieved, load factors are introduced both for qualification purposes and to account test effects.

Mathematical model correlation and validation is also obtained throughout the testing campaign, by measuring main stiffnesses during static tests and eigenfrequencies and mode shapes through modal testing.

The test campaign intends to be very comprehensive, including acceptance and qualification tests as well as measurements of intrinsic properties of materials, components and systems. Functional evaluation of systems is also included in the acceptance tests. The static tests performed during qualification analyse the behaviour of the structure and validate the mathematical models. Modal analysis is also employed to validate the models and assess stability of manufacturing processes.

### *4.2. TEST EQUIPMENT*

The test equipment encompasses all the hardware and software needed to perform the tests. The most relevant equipment designed are highlighted here.

#### *4.2.1. Test rig – Boundary conditions*

The test rig main design driver was modularity. The structure is based on singular pieces of similar shape which provide many possibilities for assembly. Thus, there is more than one structure for tests, there are many and the test structure design will depend on the test to be performed.

The modular concept enables the use of the same elements for the test campaign of the PF (Fig. 10) as well the KST. One example of the modularity application is the modal test of various single parts. By combining a few parts in a geometrically adequate form, for each part, boundary conditions are easily set.

The boundary conditions for qualification/acceptance test campaigns may be set by two different approaches, similar to flight conditions or rigid interface (relative to the test part). Similar flight conditions are complex to achieve and usually require flight/engineering of other structures, whereas a rigid interface is more straightforward, leading to the main structural requirement of the test structure – provide a very rigid interface for every test considered. The design of the test structure elements was based, at a first stage, on this requirement.

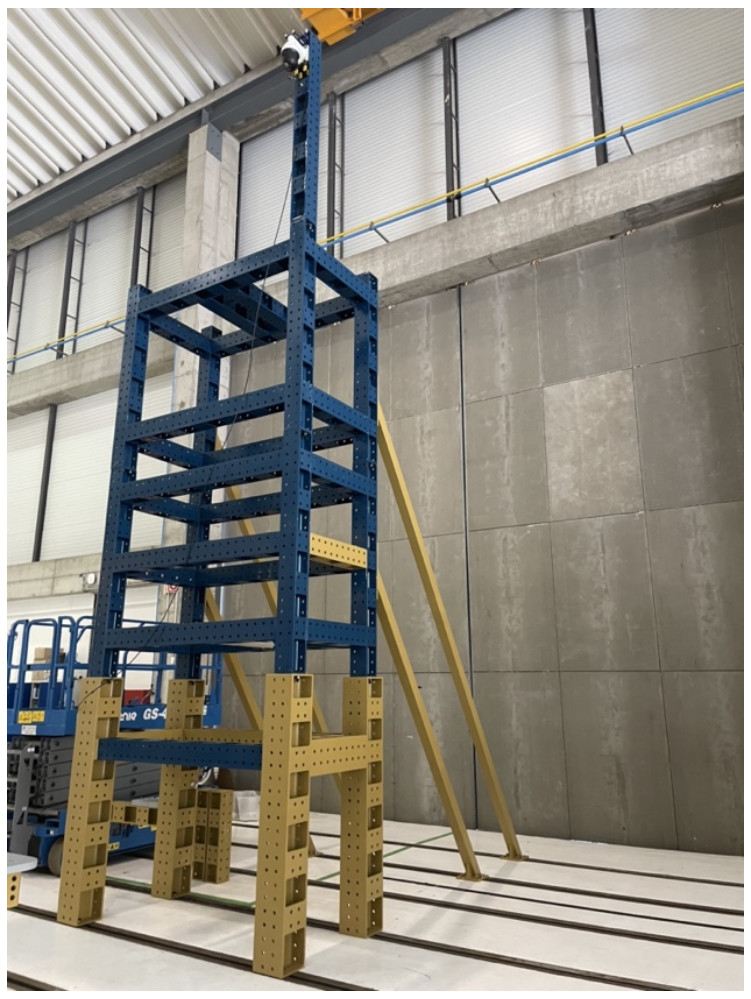

Fig. 10. Test support structure

*4.2.2. Load application devices*

Static loads are applied by means of hydraulic actuators on a close-loop control system, using precise and adequate load cells, which is able to very precisely control the force applied throughout the test, meaning very small increments of forces can be prescribed between each load increment.

Modal analysis is performed with an impact hammer on simple single components/system, while for more involved and complex systems a vibration table is used.

*4.2.3. Measuring devices*

Measurements throughout the test campaign are performed via adequate linear variable differential transformer, strain gauges, load cells and accelerometers.

*4.3. TEST CAMPAIGN*

*4.3.1. Loads*

As previously stated, test configurations apply simplified load cases to the test models which emulate flight level loads in specified locations along with test factors.

The loads applied to the PF and KST during flight have different sources, thus their tests are different and have different goals.

The PF test campaign is mainly focused on quasi-static loads, where on the KST quasi-static and dynamic environments are driving factors of the test campaign.

Quasi-static tests are performed by applying the forces in incremental steps, the load range is increased up to previously defined test levels and each step will provide information to validate the numerical models. Harmonic modal tests are performed with adequate vibration table and prescribed sine sweep. Impact modal tests are performed via impact hammer applying on successive locations, of the part, a prescribed impact and applying the roving hammer technique.

*4.3.2. Results*

The modal survey is performed to identify the global modes and provides validation for the assumptions applied on the dynamic models.

A modal test was performed on an ESPA tower, considering on free-free conditions, Fig. 11. The natural frequencies research was focused on the first five natural modes of the component. Table 4 provides a comparison between the predicted and measured values, showing agreeable values for most frequencies. Mode 2 and 4 higher differentials is mainly due to these modes having symmetric identical modes nearby, easily distinguishable in numerical models results whereas on measurements they overlap each other. The differential is acceptable and can be attributed to modelling simplifications, thus, validating the mathematical model.

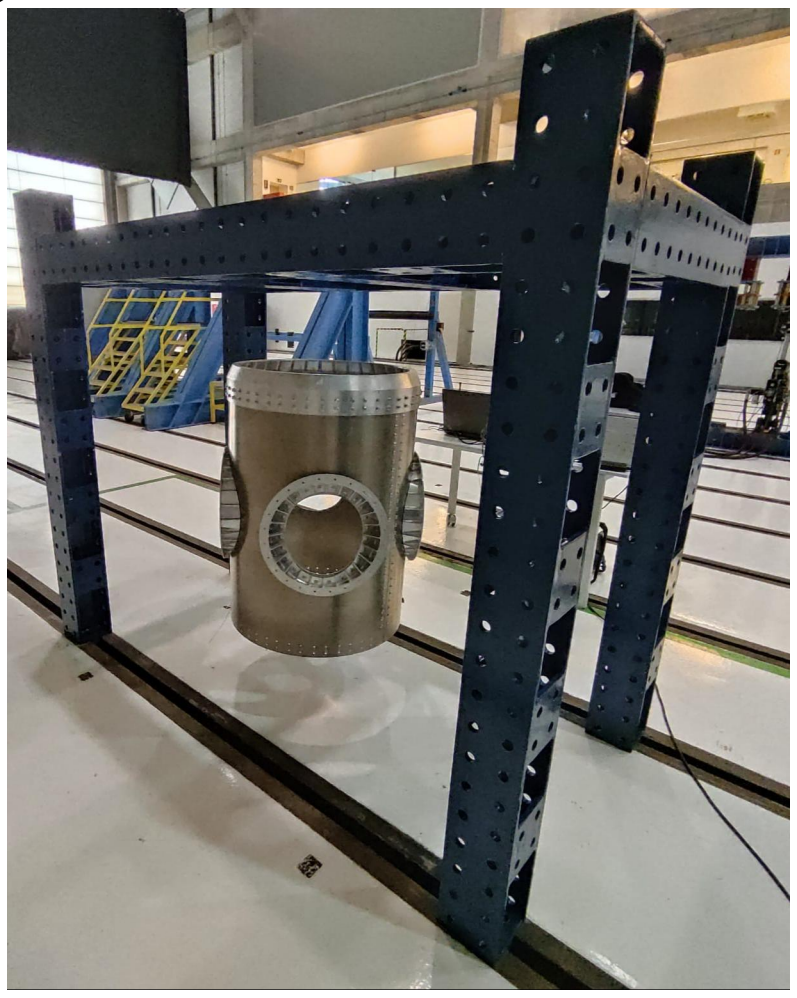

Fig. 11. Kick-Stage ESPA tower under natural frequency assessment

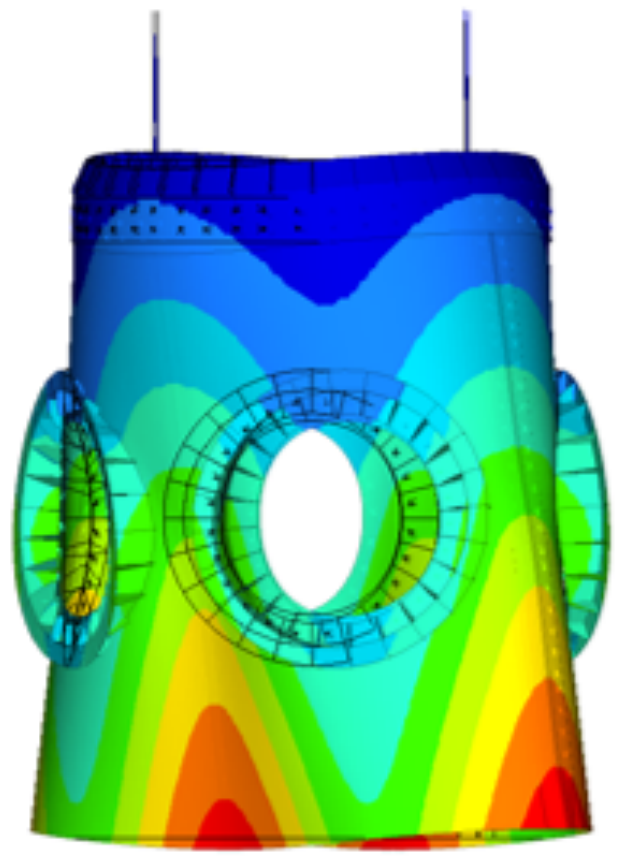

Fig. 12. ESPA tower's first natural mode

Table 4. Comparison between measured and modelled natural frequencies – ESPA Tower

| *Natural Mode* | **Differential measured/predicted [%]** |
|---|---|
| *1* | 4.3 |
| *2* | 13.7 |
| *3* | 9.7 |
| *4* | 13.7 |
| *5* | 4.7 |

A modal survey of the PF model was also performed. The results are presented on Table 5, with the first one presented in Fig. 13. The measured values of the first three natural modes are higher than the ones predicted by numerical models. Higher modes are also predicted with very reasonable accuracy. These results are very reassuring and validate the complex mathematical model of the PF. The stiffness difference measured on the test model is a result of conservative assumptions, knockdown factors and manufacturing details not easily replicated on numerical models.

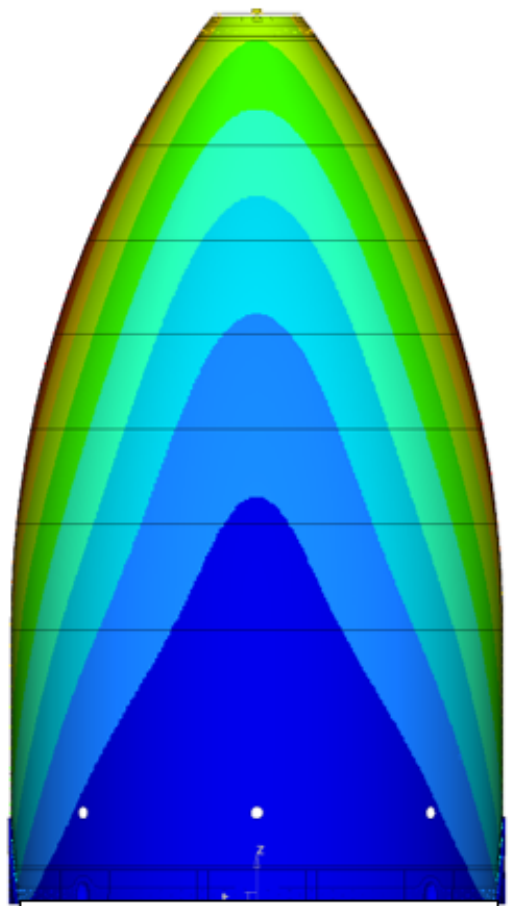

Fig. 13. Payload Fairing's first natural mode

Table 5. Comparison between measured and modelled natural frequencies – Payload Fairing

| *Natural Mode* | **Differential measured/predicted [%]** |
|---|---|
| *1* | -23.2 |
| *2* | -14.0 |
| *3* | -7.7 |
| *4* | 6.7 |
| *5* | 7.6 |

## 5. CONCLUSIONS

The current work presents the implementation of simple and well proven methodologies in sectors as the aeronautical and automotive, adapted for the New Space reality, towards cost-effective and commercially appealing small launcher space segment solutions.

These methodologies were applied in the Magellan project, which aims to develop and prototype a set of primary structures for small launchers, namely the Payload Fairing, the Kick-Stage, and the Common Separation Flanges, all to be flight tested on the RFA-One small launcher.

In Magellan, a Systems Engineering approach was implemented, more specifically with a requirements definition and validation process, a complete risk analysis throughout the entire project duration as well as a FMEA, for failure identification, control and mitigation. The enforcement of such tools and their frequent review and update with the involvement of all the stakeholders and technical teams, allowed to identify early in the project the most critical failure points. These were then mapped and tightly controlled, with corrective actions defined for the most impactful failures and identified as potential technical risks for the project. Additionally, other non-technical risks were raised and merged into a risk matrix, and respective risk exposure calculated. This approach allowed to increase the overall project efficiency in terms of time and costs, as development iterations were reduced due to early error detection. Production costs were also optimised as a first functional prototype was achieved.

Also, a brief synthesis of the procedure for verification of dimensioning applied throughout the project is presented. For both primary and secondary structures tests and models are complementary. The verification approach relies on test configurations which must be justified compared to the dimensioning configuration, from which correction factors are defined. Additionally, testing provides proof of design margins, and it is shown that it allows to validate models. A modal survey of an ESPA tower of the Kick-Stage showed frequencies similar to the predicted, thus, validating its dynamic model. The results of the modal test of the Payload Fairing also validated the model by providing a similar behaviour to the predicted. The differential value of the natural frequencies obtained is explained by conservative approaches, on the model, assumed at the beginning of the project and manufacturing details.

**Acknowledgements**

The project Magellan - Development of Micro Launcher Structures in Portugal (reference POCI-01-0247-FEDER-072246) leading to this work is co-financed by Compete 2020, Portugal 2020 and the European Regional Development Fund.